\documentclass[groupedaddress]{revtex4}

\usepackage{graphicx}

\begin{document}

\title{
Low-degree mantle convection with strongly temperature- and depth-dependent viscosity 
in a three-dimensional spherical shell
}

\author{Masaki Yoshida}
\author{Akira Kageyama}
\affiliation{
Earth Simulator Center, \\
Japan Agency for Marine-Earth Science and Technology,
Yokohama 236-0001, Japan.
}

\begin{abstract}
A series of numerical simulations of thermal convection of Boussinesq fluid 
with infinite Prandtl number, 
with Rayleigh number $10^7$, 
and with the strongly temperature- and depth- dependent viscosity 
in a three-dimensional spherical shell is carried out to study the mantle convection 
of single-plate terrestrial planets like Venus or Mars 
without an Earth-like plate tectonics.  
The strongly temperature-dependent viscosity (the viscosity contrast across the shell 
is $\geq 10^5$) make the convection under stagnant-lid short-wavelength structures. 
Numerous, cylindrical upwelling plumes are 
developed because of the secondary downwelling plumes arising from the bottom of lid.
This convection pattern is inconsistent with that inferred from 
the geodesic observation of the Venus or Mars. 
Additional effect of the stratified viscosity at the upper/lower mantle 
(the viscosity contrast is varied from 30 to 300) are investigated. 
It is found that the combination of the strongly temperature- and depth-dependent viscosity 
causes long-wavelength structures of convection 
in which the spherical harmonic degree $\ell$ is dominant at 1--4.  
The geoid anomaly calculated by the simulated convections shows a long-wavelength structure, 
which is compared with observations. 
The degree-one ($\ell = 1$) convection like the Martian mantle 
is realized in the wide range of viscosity contrast from 30 to 100 
when the viscosity is continuously increased with depth at the lower mantle.

\end{abstract}

\maketitle

\section{Introduction}
Single-plate terrestrial planets like Venus and Mars 
without an Earth-like plate tectonics 
is covered by a thick immobile lithosphere, or cold stiff lid. 
It is inferred from 
the geodesic observations (topography and gravity) of 
Venus [{\it Rappaport et al.}, 1999; {\it Konopliv et al.}, 1999] 
and Mars [{\it Smith et al.}, 1999a; 1999b] 
that the spatial structure of the thermal convection under the lid has  
relatively long-wavelength 
in which the spherical harmonic degree is dominant at $\ell=$ 2--3 or lower  
[e.g., {\it Schubert et al.}, 1990; 1997]. 
In particular, as for the Mars,  
it is generally accepted that the Martian crustal dichotomy 
was caused by a convection system dominated 
by $\ell = 1$ [e.g., {\it Sleep}, 1994; {\it Zhong and Zuber}, 2001].

In numerical simulation of mantle convection 
in the three-dimensional (3-D) Cartesian box geometry with wide aspect ratios 
[{\it Tackley}, 1996a; {\it Ratcliff et al.}, 1997, {\it Trompert and Hansen}, 1998] 
and in the spherical shell geometry [{\it Ratcliff et al.}, 1996; 1997], 
it is shown that a highly viscous lid is formed 
when the temperature-dependent viscosity is included in their models  
with the stress-free boundary condition on the top surface. 
As the viscosity contrast goes up to $10^4$--$10^5$, 
an immobile highly viscous layer (stagnant-lid) is formed.  
The convection under the stagnant-lid is characterized 
by numerous, small-scale cylindrical plumes surrounded by sheet-like downwellings 
[{\it Ratcliff et al.}, 1996; 1997; {\it Reese et al.}, 1999]. 
These convection patterns with high-degree modes 
are apparently inconsistent with the observations.  

Here we explore the possibility that the low-degree convection under a stagnant-lid 
is induced by the depth-dependent viscosity due to the higher viscous lower mantle.   
The dynamical effects of a stratified viscosity profile on 
the mantle convection without lateral viscosity variations have been studied 
by the two-dimensional (2-D) or 3-D Cartesian [e.g., {\it Hansen et al.}, 1993; {\it Tackley}, 1996b] 
and by the spherical shell 
[{\it Zhang and Yuen}, 1995; {\it Bunge et al.}, 1996; {\it Zhong et al}., 2000b] models. 
{\it Bunge et al.} [1996] have shown that a modest increase in the mantle viscosity with depth 
has a remarkable effect on the convection pattern,  
resulting in a long-wavelength structure. 
However, another important factor for the mantle viscosity, i.e., 
the strong dependence on temperature, was absent in their models. 
The purpose of this paper is to investigate the combined effects of 
(i) depth-dependence and (ii) strong temperature-dependence on the viscosity 
in the resulting convection pattern. 

\section{Simulation Model}

The mantle convection is numerically treated as a thermal convection 
in a 3-D spherical shell of a Boussinesq fluid with infinite Prandtl number heated from the bottom boundary. 
The aspect ratio of the spherical shell $\hat{r}_0/\hat{r}_1$ is 0.55, 
which is a characteristic value of the terrestrial planets, 
where $\hat{r}_0$ and $\hat{r}_1$ are the radii of the inner and outer spheres, respectively. 
Equations of mass, momentum, and energy conservation governing
the mantle convection are scaled to a non-dimensional form as follows [e.g., {\it Schubert et al.}, 2001],  
\begin{equation}
\mathbf{\nabla} \cdot \mathbf{v} = 0,
\end{equation}
\begin{equation}
- \mathbf{\nabla} p 
+ \mathbf{\nabla} \cdot 
\left\{ \eta 
 \left( \mathbf{\nabla} \mathbf{v} + \mathbf{\nabla} \mathbf{v}^{tr} \right) 
\right\} + Ra_r T \mathbf{e}_r = 0, 
\end{equation}
\begin{equation}
\frac{\partial T}{\partial t} 
+ \mathbf{v} \cdot \mathbf{\nabla} T
= \mathbf{\nabla}^2 T + H_r, 
\end{equation}
where $\mathbf{v}$ is the velocity vector, $p$ pressure, 
$T$ temperature, $t$ time, and $\mathbf{e}_r$ is the unit vector 
in the $r$-direction. 
The superscript $tr$ indicates the tensor transpose. 
The Rayleigh number $Ra$ scaled by the thickness of the spherical shell $\hat{D}$ is given by, 
\begin{equation}  
Ra \equiv Ra_r \left( \frac{\hat{D}}{\hat{r_1}} \right)^3 
   = \frac{\hat{\rho} \hat{g} \hat{\alpha} \Delta \hat{T} \hat{D}^3}{\hat{\kappa} \hat{\eta}_{ref}}, 
\end{equation}
where  
$\hat{\rho}$ is the density, $\hat{g}$ gravitational acceleration, 
$\hat{\alpha}$ thermal expansivity, 
$\Delta \hat{T} (= \hat{T}_{bot} - \hat{T}_{top}$) the temperature difference 
between the bottom temperature $\hat{T}_{bot}$ on the inner sphere 
and the top temperature $\hat{T}_{top}$ on the outer sphere, 
$\hat{\kappa}$ thermal diffusivity,  
and $\hat{\eta}_{ref}$ is the reference viscosity (see equation~(6) below). 
The hats stand for dimensional quantities.  

The internal heating rate $H$ scaled by the thickness of the spherical shell $\hat{D}$ is given by,
\begin{equation}  
H \equiv H_r \left( \frac{\hat{D}}{\hat{r_1}} \right)^2 
    = \frac{\hat{Q} \hat{D}^2}{\hat{\kappa} \hat{c}_p \Delta \hat{T}}, 
\end{equation}
where $\hat{Q}$ is the internal heating rate per unit mass, 
and $\hat{c}_p$ is the specific heat at constant pressure. 
In this study, in order to focus on the effects of the temperature- and depth-dependent viscosity, 
all the material properties other than viscosity (such as thermal expansivity and thermal diffusivity) 
are assumed to be constant. 
The viscosity $\eta$ depends on the temperature $T$ and depth $d$ as 
\begin{equation}
\eta(T, d) = \eta_{ref} (d) \exp \left[ -E \left( T - T_{ref} \right) \right], 
\end{equation}
where $\eta_{ref} (d)$ is the viscosity at the reference temperature $T = T_{ref}$. 
The non-dimensional ``activation parameter'' $E$ 
represents the degree of viscosity contrast between the top and bottom surfaces. 
The velocity boundary at the top and bottom surfaces of the spherical shell 
are given by impermeable and the stress-free conditions. 
The boundary conditions for $T$ at the top and bottom surfaces are given by 
$T_{bot} = 1$ and $T_{top} = 0$. 

The basic equations (1)--(3) are solved by a second-order finite difference discretization.   
A kind of the overset (Chimera) grid system, Yin-Yang grid [{\it Kageyama and Sato}, 2004],  
is used for the computational grid (Figure~1). 
The Yin-Yang grid is composed of two component grids (Yin grid and Yang grid) 
that have exactly the same shape and size (Figure~1a). 
A component grid of the Yin-Yang grid is a low-latitude part 
of the usual latitude-longitude grid on the spherical polar coordinates. 
The Yin-Yang grid is suitable to solve 
the mantle convection problems because it automatically avoids the pole problems, 
i.e., the coordinate singularity and grid convergence 
that are inevitable in the usual latitude-longitude grid (Figure~1b). 
Following the general overset grid method,  
the data on the boarder of each component grid are matched by mutual interpolation. 
All the basic quantities---$ \mathbf{v}$, $p$, $T$, and $\eta$---are spatially discretized 
and located in the same grid points (collocation grid method). 
The details of the Yin-Yang grid can be found in {\it Kageyama and Sato} [2004]. 
See our previous paper [{\it Yoshida and Kageyama}, 2004] for 
its application to the mantle convection with detailed benchmark and validation tests. 

The grid points in each component grid are 
$102~\times~54~\times~158$ (in $r$-, $\theta$-, and $\phi$-directions). 
Thus the total grid size for a whole spherical shell is 
$102~\times~54~\times~158~\times~2$ (for Yin and Yang grids). 
The convergences of the solutions were confirmed by 
changing the numerical resolution with $66~\times~33~\times~104~\times~2$. 
Time development of the convections are calculated until 
averaged quantities, such as Nusselt number and root-mean-square velocity, become stationary.

\section{Results}

Calculations carried out in this paper are summarized in Table. 1.

\subsection{Constant viscosity and only temperature-dependent viscosity convections}

Before we go into details of the combined effects of 
the temperature- and depth-dependent viscosity on the convection, 
we study the phenomenology of convection pattern changes caused only 
by the temperature-dependent viscosity.

Figure~2a shows a snapshot of the residual temperature of Case~{\tt r7e0} in Table 1, 
in which the viscosity is constant, i.e., $E=0$ in equation~(6). 
The Rayleigh number $Ra$ is $10^7$, 
which is about one order of magnitude smaller than the value of terrestrial planets. 
(Later, we will show that the convective pattern is unchanged even 
when the Rayleigh number is increased 10$^8$ in Case {\tt r8e6w2}.)  
The thermal structure of Figure~2a shows a typical pattern observed in the 3-D spherical shell geometry.  
The convective flow is composed of narrow, cylindrical upwelling (hot) plumes surrounded by 
a network of long downwelling (cold) sheets. 
This structure is common for the purely bottom-heated convection. 
To analyze the spatial structure, the power spectrum by the spherical harmonics $Y_{\ell}^m$
of temperature field is plotted in the right panels of Figure~2. 
The small scale structure ($\ell \geq 10$) is dominant in the middle depth, 
and the large scale structure ($\ell \leq 6$) is dominant near the top and bottom surfaces 
that is associated with the thermal boundary layers. 
The radial profile of horizontally averaged temperature is shown in Figure~3a. 
The volume-averaged temperature is 0.26 in this case.  
As we expect, compared with the convections with high Rayleigh number 
and the strong internal heating 
[{\it Bunge et al.}, 1996; {\it Yoshida et al.}, 1999; {\it Zhong et al.}, 2000b], 
the thermal structure of this purely bottom-heated convection 
are dominated by considerably long-wavelength structure. 

Figure~2b shows the results of Case~{\tt r7e6r} 
where the reference temperature  $T_{ref} = 0.5$.  
The activation parameter is taken to be $E = \ln (10^6) = 13.8155$. 
The spectrum of temperature field of Case~{\tt r7e6r} (right panel of Figure~2b)  
shows that the power is concentrated around $\ell = 6$--$10$ throughout the depth 
that is associated with convecting cells under the stagnant-lid.
For this case, the volume averaged temperature is 0.72, 
which is larger than the constant viscosity convection (Figure~3b).  

In our previous paper [{\it Yoshida and Kageyama}, 2004],  
we did not report the cases 
when the viscosity of mantle materials has a strong temperature dependence.   
The regime of the flow state under the strong temperature-dependent viscosity 
in the spherical shell convection was examined by {\it Ratcliff et al.} [1996; 1997]. 
For the comparison with the previous works, 
the reference temperature $T_{ref}$ in equation~(6) is fixed 
to the bottom temperature $T_{bot}$ in the followings. 
Therefore, the viscosity $\eta_{ref}$ is now the viscosity at the bottom. 
The viscosity contrast across the spherical shell is defined by 
$\gamma_{\eta} \equiv \eta(T_{top}) / \eta(T_{bot}) = \exp(E)$. 

Shown in Figure~4a is a regime diagram for convective flow pattern. 
Approximate regime boundaries are drawn.  
Our simulation results for $Ra_{bot} = 10^{6}$--$10^{7}$ are shown in this diagram. 
The previous results by {\it Ratcliff et al.} [1997] (3-D Cartesian and spherical shell models) 
and {\it Trompert and Hansen} [1998] (3-D Cartesian model) 
are also included in the diagram. 
Our results basically support the previous results by {\it Ratcliff et al.} [1996; 1997]:  
The convecting pattern is classified into three regimes  
defined by {\it Solomatov} [1995]  in the order of increasing $\gamma_{\eta}$;  
the ``mobile-lid'' regime (Figure~4b); the ``sluggish-lid'' regime (Figure~4c);  
and the ``stagnant-lid'' regime (Figure~4d).

The moderate viscosity contrast ($\gamma_{\eta} = 10^3$--$10^4$) 
produces the large-scale convection, or the sluggish-lid regime. 
In our previous paper [{\it Yoshida and Kageyama}, 2004], 
we showed that the convection at $Ra_{bot} = 10^6$ and $\gamma_{\eta} = 10^4$ (Case~{\tt r6e4}) 
has a two cell pattern that consists of one downwelling and two cylindrical upwellings (Figure~4b) 
[{\it Ratcliff et al.}, 1995; 1996; 1997, {\it Zhong et al}., 2000b; 
{\it Yoshida and Kageyama}, 2004; {\it Stemmer et al.}, 2004; {\it McNamara and Zhong}, 2005a]. 
In contrast, at $Ra_{bot} = 10^7$ and $\gamma_{\eta} = 10^4$ (Case~{\tt r7e4}), 
the convection pattern comes to have the degree-one pattern; the one cell structure that 
consists of a pair of cylindrical downwelling plume and 
cylindrical upwelling plume (Figure~4c). 
This indicates that the convecting structure in the sluggish-lid regime is sensible to the Rayleigh number.

The convective flow pattern that belongs to the stagnant-lid regime emerges 
when $\gamma_{\eta} \geq 10^5$. 
The stagnant-lid prevents the heat flux through the top boundary and 
leads to a small temperature difference in the mantle below the lid. 
The characteristic horizontal thermal structure has short wavelengths comparable to 
the thickness of the mantle (Figures~4d and 4e). 
This convective pattern in the stagnant-lid regime is also observed in the previous results 
in a 3-D spherical shell geometry [e.g., {\it Reese et al.}, 1999b]. 
This convective feature would be caused by the secondary downwelling plumes 
leaving from the base of stagnant-lid. 
At $\gamma_{\eta} \geq 10^6$ (Cases~{\tt r7e6}), the connected network of sheet-like downwelling 
reaches to the mid depth of convecting layer (Figure 4d). 
When $\gamma_{\eta}$ is further increased to $10^8$ (Cases~{\tt r7e8} and {\tt r7eA}),  
the stagnant-lid become rather thick, and we clearly observe large, 
mushroom-shaped upwelling plumes (Figure~4e).

\subsection{Both temperature- and depth-dependent viscosity convections}

To investigate the transition in the convective pattern by adding 
the depth-dependent viscosity (a viscosity stratification),  
we investigate two kinds of viscosity profiles. 
First we examine cases in which 
the viscosity jumps at the phase transition boundary between the upper and lower mantle.  
Second we examine cases in which the viscosity 
smoothly increases with depth in the lower mantle. 
The ratio of thickness between lower and upper mantle, $d_L/d_U$, is 3.39, 
comparable to that in Earth's mantle. 
Since the actual viscosity contrast  in the depth of the terrestrial planets is not fully constrained, 
we take it as a parameter within a plausible range between $10^{1.5}$ ($\approx 30$) 
$\leq$ $\eta_L/ \eta_{ref}$ $\leq$ $10^{2.5}$ ($\approx 300$) 
[e.g., {\it Davies and Richards}, 1992; {\it Karato}, 2003] 
where $\eta_L$ is the viscosity of lower mantle.  
In six cases (Cases~{\tt r7e6v1} to {\tt r7e6v3}, and Cases~{\tt r7e6w1} to {\tt r7e6w3}), 
the initial condition is taken from the stationary state of Case~{\tt r7e6r}, shown in Figure~2b.
The reference temperature $T_{ref}$ in equation~(6) is fixed at 0.5.  
The Rayleigh number defined by $\eta_{ref}$ is fixed at 10$^{7}$.  

Shown in Figure~5 are the results of three cases (Cases~{\tt r7e6v1}, {\tt r7e6v2}, and {\tt r7e6v3}) 
in which the viscosity jumps at the upper/lower mantle boundary. 
Figure~5a shows a snapshot of the residual temperature of Case~{\tt r7e6v1}   
with $\eta_L/\eta_{ref} = 10^{1.5}$. 
Compared with the convection in which the viscosity depends only on the temperature (Figure~2b), 
we find that the convective flow pattern obviously 
has longer length scale. 
The thermal spectrum indicates a shift to smaller degrees, 
and the peak is located between $\ell = 2$ and $\approx 10$. 
As $\eta_L/\eta_{ref}$ is further increased 
(Cases~{\tt r7e6v2} shown in Figure 5b and {\tt r7e6v3} in Figure 5c), 
the thermal structure significantly shifts to lower modes. 
The power spectrum shows a concentration in $\ell \le 6$  
with the peak of $\ell = 2$--$4$ for $\eta_L/\eta_{ref} = 10^{2.0}$ (Figure~5b), 
and $\ell = 2$--$3$ for $\eta_L/\eta_{ref} = 10^{2.5}$ (Figure~5c) throughout the depth. 
As the amount of the viscosity jump $\eta_L/\eta_{ref}$ increases, 
the temperature drop in the bottom thermal boundary layer grows, 
which leads to the lower internal temperature of the mantle (Figure~6). 

To see this spatial scale change of the convection caused 
by the depth-dependent viscosity in more detail, 
we analyzed the time sequence of 
the Nusselt number, the root-mean-square velocity averaged over the entire mantle, 
and the peak mode at each depth 
for the Case~{\tt r7e6v3} with $\eta_L/\eta_{ref} = 10^{2.5}$. 
At the initial stage of the simulation run, the convection is dominated by $\ell = 7$--$9$ modes 
throughout the depth which reflects the initial condition (Figure~2b). 
As time goes on, the convective flow reaches to a saturated state (Figure~7a), 
and the low-degree component develops from the upper part to middle part of mantle; 
the peak mode shifts from $\ell = 9$ to $3$ there (Figure~7b). 
This indicates that the stagnant-lid is broken, and then, the convection cells 
are re-organized into the convection state with the low modes.  

To compare with the observation, 
we have calculated the geoid anomaly for Cases~{\tt r7e6v1} to {\tt r7e6v3}. 
We followed the method of the calculation of the geoid anomaly 
described in {\it Hager and Clayton} [1989]. 
The physical parameters used in the calculation are set to those possibly relevant to Venus (Table 2). 
Figure~8 shows the distribution of calculated geoid anomaly  
where $\eta_{L}/\eta_{ref}$ is (a) 1 (i.e., no viscosity stratification), 
(b) $10^{1.5}$, (c) $10^{2.0}$, and (d) $10^{2.5}$.  
The results are shown by the spherical harmonics modes up to $\ell=24$. 
Figure~8e shows the power spectrum for each case. 
The mode amplitude with the viscosity stratification peaks at $\ell = 2$--$4$. 
When the stratified viscosity is absent (Figure~8a), 
$\ell = 5$--$10$ modes are strong (see the arrow in Figure~8e).  
On the other hand, as $\eta_{L}/\eta_{ref}$ increases (Figures~8c and 8d),  
the power spectrum peaks at $\ell = 2$ and the higher degree components ($\ell \ge 10$) 
are remarkably decreased.   
This is consistent with the spectrum constructed from 
the observed geoid anomaly of the Venus [{\it Konopliv et al.}, 1999] (Figure~8e). 

Next, we investigate the cases with smoothly increased viscosity with depth rather than the jump. 
In three cases (Cases~{\tt r7e6w1}, {\tt r7e6w2}, and {\tt r7e6w3}) shown in Figure~9,  
the viscosity contrast between the upper/lower mantle boundary and 
the bottom of mantle is $\Delta \eta_L =$ (a) $10^{1.5}$, (b) $10^{2}$, and (c) $10^{2.5}$. 
The initial condition is again the state shown in Figure~2b. 
We see from both the residual temperature and the spectrum 
that the dominant power is concentrated on the smaller degrees in all the cases. 
At $\Delta \eta_L = 10^{1.5}$--$10^{2}$, 
the peak is located at $\ell = 1$, or one-cell convection (Figures~9a and 9b). 
On the other hand, at $\Delta \eta_L = 10^{2.5}$, the peak is located at $\ell = 2$,  
or the two-cell convection (Figure~9c). 
The horizontally averaged temperature and viscosity profile is shown in Figure~10. 
Note that the viscosity contrast in the lid are almost identical among the three cases 
(see the arrow in the right panel of Figure~10).  
This suggests that the transition between degree-one and degree-two convection  
is sensitive to the magnitude of the viscosity stratification. 
We found that the patterns (degree-one or degree-two) are not affected by the increase of $E$ 
up to 8 (Case~{\tt r7e8w2}) or to 10 (Case~{\tt r7eAw2}). 
This pattern is also unchanged when the internal heating is included ($H=20$) (Case~{\tt r7e6w2h}), 
or the Rayleigh number is increased to 10$^{8}$ (Case~{\tt r8e6w2}). 
The patterns $\ell = 1$ or $2$ are mainly controlled by the viscosity contrast $\Delta \eta_L$.

\section{Conclusions and Discussion}

The convection with strongly temperature dependent viscosity 
under the stress-free boundary condition has short wavelength structure 
when the depth-dependent viscosity is ignored.   
This feature is inconsistent with the convection inferred from 
the geodesic observations on the single-plate planets like Venus and Mars. 
We have found that the combination of temperature- and depth-dependent viscosity 
produces the convection with the spherical harmonics degree $\ell = 1$--$4$. 
The geoid anomaly calculated from these simulation data also generates 
large scale length, which is consistent with the observation.
{\it Schubert et al.} [1990; 1997] have shown 
that convections with rigid boundary condition on the top surface 
can lead to $\ell = 1$--$3$ structures. 
In their model, however, the viscosity of fluid is spatially constant. 
Our finding is that, by considering more realistic viscosity profiles, 
the low-degree pattern can be reproduced in the convection model  
with stress-free boundary condition on the top surface. 

The previous convection models without the temperature-dependent viscosity 
[e.g., {\it Hansen et al.}, 1993; {\it Zhang and Yuen}, 1995; {\it Bunge et al.}, 1996] 
have already produced the large scale flow pattern by considering the viscosity stratification. 
This could be explained by the enhanced value of viscosity in the lower mantle. 
In our model with strongly temperature-dependent viscosity, 
the large scale convection seems to be realized 
by the change of convecting regime, 
from the stagnant-lid regime into the sluggish-lid regime, 
which is caused by the viscosity stratification. 
A major difference between their results and ours is that 
a highly viscous lid is naturally formed on the top 
owing to the inclusion of the temperature-dependent viscosity effect in this study. 

To date, several mechanisms have been proposed for the degree-one convection of the Martian mantle.  
For example, the endothermic phase transition just above core-mantle 
boundary in Martian mantle 
with the rigid boundary condition [{\it Harder and Christensen}, 1996; {\it Breuer et al.}, 1998; {\it Harder}, 1998],  
and a priori high-viscous lid [{\it Harder}, 2000] on the top surface boundary 
without any phase transitions. 
The small core, in other words, the thicker convecting shells of the mantle 
may lead to the degree-one convection 
in the ancient Mars [{\it Schubert et al.}, 1990] and the Moon [{\it Zhong et al.}, 2000a]. 
{\it McNamara and Zhong} [2005a] have recently found that 
the internal heating plays a role in 
increasing flow wavelength and forming the degree-one convection 
in convections in which the viscosity moderately depends on temperature. 
One of our findings in this paper is that 
the degree-one convection can be relatively easily reproduced 
when both effects of the temperature- and depth-dependence on the viscosity are taken into account.
Although the degree-one convection appears 
even when the depth-dependence is absent (Figure~4c), 
the parameter range for this pattern is rather narrow; 
it is sensitive to the Rayleigh number. 
On the other hand, when the viscosity in the lower mantle is continuously increased with depth,  
the degree-one ($\ell = 1$) convection like the Martian mantle 
is realized in the wide range of viscosity contrast from 30 to 100. 

It is an interesting possibility that the transition of the convecting patterns  
between low-degree convective mode and one-degree mode took place in the planets.
We have not directly observed the transition of convecting mode in our simulations.  
The physical parameters ($Ra$ and/or $E$ in this study) 
to characterize the convective pattern are fixed in our simulations. 
However, we would like to point out again a drastic difference of the convection patterns  
between relatively close conditions: 
the convection of degree-two at $Ra = 10^6$ (Figure~4b) [{\it Yoshida and Kageyama}, 2004], and 
the convection of degree-one at $Ra = 10^7$ (Figure~4c). 
This sensitive change has not been reported so far.

Our simulation results will not be directly applied to the Earth's mantle, 
because the effects of the plate tectonics would be comparable to 
the effects of the depth-dependent viscosity,  
as proposed by {\it Bunge and Richards} [1996] and {\it Bunge et al.} [1996; 1998] 
from their model without temperature-dependent viscosity. 
Existence of a stationary continental lithosphere [{\it Yoshida et al.}, 1999], 
a drifting continental lithosphere [{\it Phillips and Bunge}, 2005],  
and plate motion on the top surface boundary [{\it Zhong et al}., 2000b] also 
transform the small scale convection patterns in high Rayleigh number convection  
into the large scale convection patterns.

\begin{acknowledgments}
The authors are grateful to two anonymous reviewers for helpful comments. 
All the simulations were carried out on the Earth Simulator 
at Japan Agency for Marine-Earth Science and Technology. 
A part of figures in this paper was produced 
using the Generic Mapping Tools (GMT) 
released by P. Wessel and W. H. F. Smith (1998). 
\end{acknowledgments}

\section*{References}

\begin{description}
\item[]
Breuer, D., D. A. Yuen, T. Spohn, and S. Zhang (1998), 
Three dimensional models of Martian convection with phase transitions, 
Geophys. Res. Lett. \textit{25}(3), 229--232. 
\item[]
Bunge, H. -P., and M. A. Richards (1996), 
The origin of long-wavelength structure in mantle convection: 
effects of plate motions and viscosity stratification, 
Geophys. Res. Lett. \textit{23}(21), 2987--2990.
\item[]
Bunge, H. -P., M. A. Richards, and J. R. Baumgardner (1996), 
Effect of depth-dependent viscosity on the planform of mantle convection, 
\textit{Nature}, \textit{379}, 436--438.
\item[]
Bunge, H. -P., M. A. Richards, C. Lithgow-Bertelloni, 
J. R. Baumgardner, S. Grand, and B. Romanowicz (1998), 
Time scales and heterogeneous structure in geodynamic earth models, 
\textit{Science}, \textit{280}, 91--95.
\item[]
Davies, G. F., and M. A. Richards (1992), 
Mantle convection, 
{\it J. Geol.}, {\it 100}, 151--206. 
\item[]
Hager, B. H., and R. W. Clayton (1989), 
Constraints on the structure of mantle convection using seismic observations, 
flow models and the geoid. 
In: Peltier, W.R. (Ed.), {\it Mantle Convection}. Gordon 
and Breach, New York, pp. 657--763.
\item[]
Hansen, U., D. A. Yuen, S. E. Kroening, and T. B. Larsen (1993), 
Dynamical consequences of depth-dependent thermal expansivity 
and viscosity on mantle circulations and thermal structure, 
{\it Phys. Earth. Planet. Inter.}, 77, 205--223.
\item[]
Harder, H. (1998), 
Phase transitions and the three-dimensional planform of thermal convection 
in the martian Mantle, 
J. Geophys. Res. \textit{103}(E7), 16775--16797. 
\item[]
Harder, H. (2000), 
Mantle convection and the dynamic geoid of Mars, 
Geophys. Res. Lett. \textit{27}(3), 301--304. 
\item[]
Harder, H., and U. R. Christensen (1996), 
A one-plume model of martian mantle convection, 
\textit{Nature}, \textit{380}, 507--509.
\item[]
Kageyama, A., and T. Sato (2004),
The ``Yin-Yang grid'': An overset grid in spherical geometry,  
\textit{Geochem.\ Geophys.\ Geosyst.},
5(9), Q09005, doi:10.1029/2004GC000734. 
\item[]
Karato, S. (2003),  
\textit{The Dynamic Structure of the Deep Earth: An Interdisciplinary Approach}, 
241 pp., Princeton University Press.
\item[]
Konopliv, A. S., W. B. Banerdt, and W. L. Sjogren (1999),
Venus gravity: 180th degree and order model,
\textit{Icarus}, \textit{139}, 3--18.
\item[]
McNamara, A. K., and S. Zhong (2005a), 
Degree-one mantle convection: Dependence on internal heating 
and temperature-dependent rheology, 
Geophys. Res. Lett. \textit{32}, L01301, doi:10.1029/2004GL021082.
\item[]
McNamara, A. K., and S. Zhong (2005b), 
Thermochemical structures beneath Africa and the Pacific Ocean, 
\textit{Nature}, \textit{437}, 1136--1139.  
\item[]
Phillips, B. R., and H-. P. Bunge (2005),  
Heterogeneity and time dependence in 3D spherical mantle convection models with continental drift, 
Earth Planet. Sci. Lett. \textit{233}, 121--135.
\item[]
Rappaport, N. J., A. S. Konopliv, and A. B. Kucinskas (1999),
An improved 360 degree and order model of Venus topography, 
\textit{Icarus}, \textit{139}, 19--31. 
\item[]
Ratcliff, J. T., G. Schubert, and A. Zebib (1995), 
Three-dimensional variable viscosity convection of an infinite 
Prandtl number Boussinesq fluid in a spherical shell,  
Geophys. Res. Lett. \textit{22}(16), 2227--2230. 
\item[]
Ratcliff, J. T., G. Schubert, and A. Zebib (1996), 
Effects of temperature-dependent viscosity on thermal convection,  
in a spherical shell, 
\textit{Physica D}, \textit{97}, 242--252.
\item[]
Ratcliff, J. T., P. J. Tackley, G. Schubert, and A. Zebib (1997),  
Transitions in thermal convection with strongly variable viscosity,
Phys. Earth Planet. Inter. \textit{102}, 201--212.
\item[]
Reese, C. C., V. S. Solomatov, J. R. Baumgardner, and W. -S. Yang (1999), 
Stagnant lid convection in a spherical shell, 
Phys. Earth Planet. Inter. \textit{116}, 1--7. 
\item[]
Schubert, G., D. Bercovici, and G. A. Glatzmaier (1990),
Mantle dynamics in Mars and Venus: Influence of an immobile lithosphere 
on three dimensional mantle convection,
J. Geophys. Res. \textit{95}(B9), 14105--14129.
\item[]
Schubert, G., D. L. Turcotte, and P. Olson (2001), 
\textit{Mantle Convection in the Earth and Planets}, 
940 pp., Cambridge Univ. Press., New York.
\item[]
Schubert, G., V. S. Solomatov, P. J. Tackley, and D. L. Turcotte (1997),
Mantle convection and the thermal evolution of Venus, 
in \textit{Venus II - Geology, Geophysics, Atmosphere, and Solar Wind Environment}, 
edited by S. W. Bougher, D. M. Hunten, R. J. Phillips, 
University of Arizona Press, Tucson, Arizona, pp. 1245--1288.

\item[]
Sleep, N. H. (1994), 
Martian plate tectonics, 
J. Geophys. Res. {\it 99}(25), 5639--5655.
\item[]
Smith, D. E., Sjogren, G. L. Tyler, G. Balmino, F. G. Lemoines, and A. S. Konopliv (1999a), 
The Gravity Field of Mars: Results from Mars Global Surveyor, 
\textit{Science}, \textit{286}, 94--97.
\item[]
Smith, D. E., M. T. Zuber, S. C. Solomon, R. J. Phillips, J. W. Head, J. B. Garvin, W. B. Banerdt, 
D. O. Muhleman, G. H. Pettengill, G. A. Neumann, F. G. Lemoine, J. B. Abshire, O. Aharonson, 
C. D. Brown, S. A. Hauck, A. B. Ivanov, P. J. McGovern, H. J. Zwally, T. C. Duxbury (1999b), 
The global topography of Mars and implications for surface evolution,  
\textit{Science}, \textit{284}, 1495--1503.
\item[]
Solomatov, V. S. (1995), 
Scaling of temperature- and stress-dependent viscosity convection, 
\textit{Phys. Fluids}, \textit{7}(2), 266--274.
\item[]
Solomatov, V. S., and L. -N. Moresi (1996),
Stagnant lid convection on Venus, 
J. Geophys. Res. \textit{101}(E2), 4737--4753. 
\item[]
Stemmer, K., H. Harder, and U. Hansen (2004),  
Thermal convection in a 3D spherical shell with strongly temperature and pressure dependent, 
\textit{Eos Trans. AGU},  \textit{85}(47), Fall Meet. Suppl., Abstract T11E-1331. 
\item[]
Tackley, P. J. (1996a), 
Effects of strongly variable viscosity on three-dimensional 
compressible convection in planetary mantles, 
J. Geophys. Res. \textit{101}(B2), 3311--3332. 
\item[]
Tackley, P. J. (1996b), 
On the ability of phase transitions and viscosity layering to induce long wavelength heterogeneity in the mantle, 
Geophys. Res. Lett. \textit{23}(15), 1985--1988.
\item[]
Turcotte, D. L., and G. Schubert (2002), 
\textit{Geodynamics}, 2nd. ed., pp. 456, Cambridge Univ. Press, U.K.
\item[]
Trompert, R. A., and U. Hansen, U. (1998),  
On the Rayleigh number dependence of convection with a strongly temperature-dependent viscosity, 
\textit{Phys.\ Fluids}, \textit{10}, 351--360.

\item[]
Wessel, P., and W. H. F. Smith (1998), 
New, improved version of the Generic Mapping Tools released, 
{\it EOS. Trans. AGU}, 79, 579.

\item[]
Yoshida, M., and A. Kageyama (2004), 
Application of the Yin-Yang grid to a thermal convection of 
a Boussinesq fluid with infinite Prandtl number in a three-dimensional spherical shell, 
Geophys. Res. Lett. \textit{31}(12), L12609, doi:10.1029/2004GL019970.
\item[]
Yoshida, M., Y. Iwase, and S. Honda (1999), 
Generation of plumes under a localized high viscosity lid 
on 3-D spherical shell convection, 
Geophys. Res. Lett. \textit{26}(7), 947--950.
\item[]
Zhang, S., and D. A. Yuen (1995), 
The influences of lower mantle viscosity stratification 
on 3D spherical-shell mantle convection, 
Earth Planet. Sci. Lett. \textit{132}, 157--166. 
\item[]
Zhong, S., and M. T. Zuber (2001),
Degree-1 mantle convection and Martian crustal dichotomy, 
Earth Planet. Sci. Lett. \textit{189}, 75--84.
\item[]
Zhong, S., E. M. Parmentier, and M. T. Zuber (2000a), 
A dynamic origin for the global asymmetry of lunar mare basalts, 
Earth Planet. Sci. Lett. \textit{177}, 131--140.
\item[]
Zhong, S., M. T. Zuber, L. Moresi, and M. Gurnis (2000b), 
Role of temperature-dependent viscosity and surface plates 
in spherical shell models of mantle convection, 
J. Geophys. Res. \textit{105}(B5), 11063--11082.
\end{description}

\begin{figure}
\noindent\includegraphics[width=15pc]{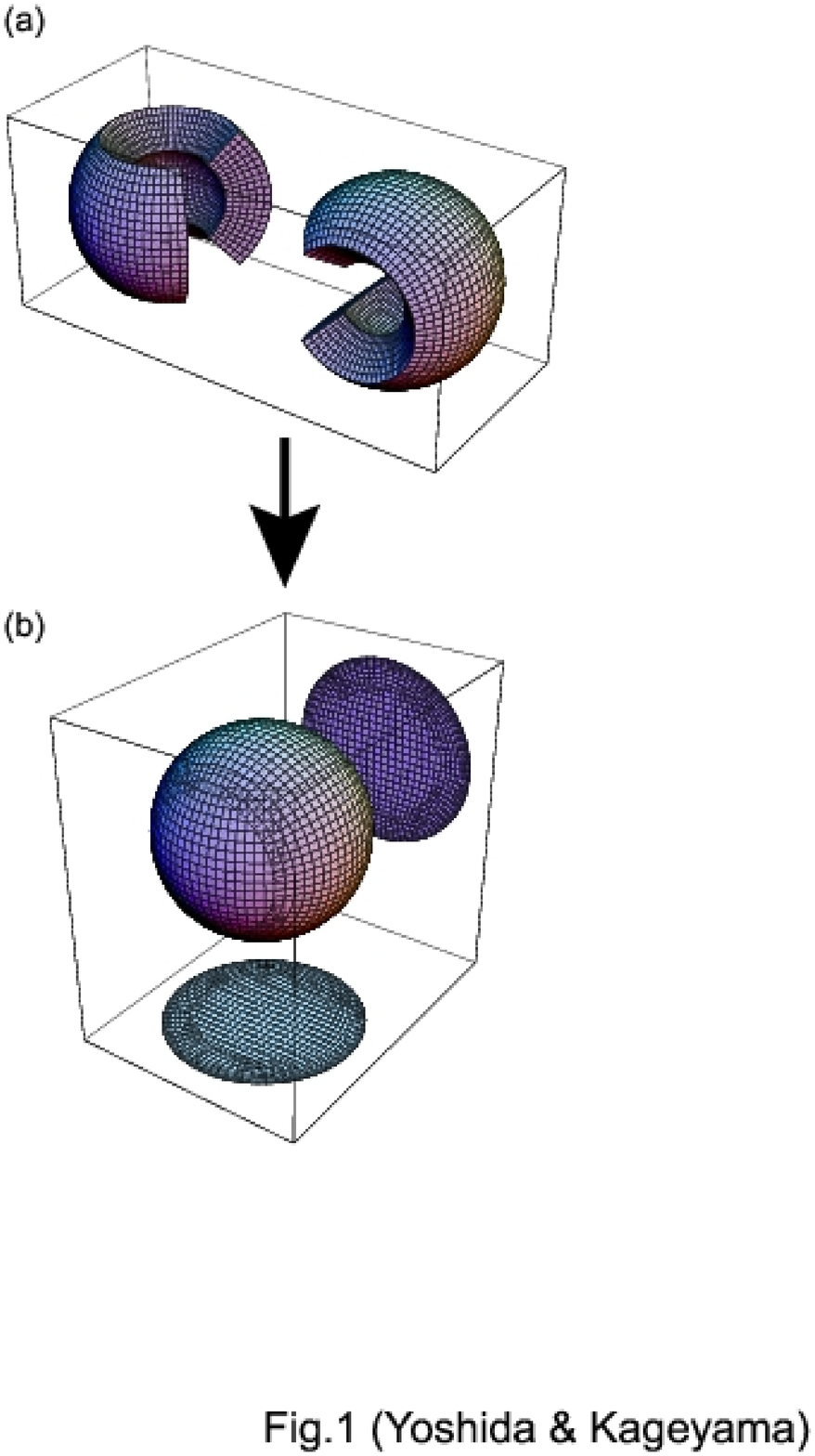}
\caption{
The Yin-Yang grid. 
Two component grids of the Yin-Yang grid are identical (the same shape and size):  
(a) The low latitude part $(\pi/4 \le \theta \le 3\pi/4, -3\pi/4 \le \phi \le 3\pi/4)$ 
of the latitude-longitude grid.  
(b) They partially overlap each other on their boarders to cover a spherical surface in pair. 
As it is apparent, the Yin-Yang grid has neither a coordinate singularity, nor grid convergence; 
the grid spacings are quasi-uniforms on the sphere. 
}
\end{figure}

\begin{figure*}
\noindent\includegraphics[width=20pc]{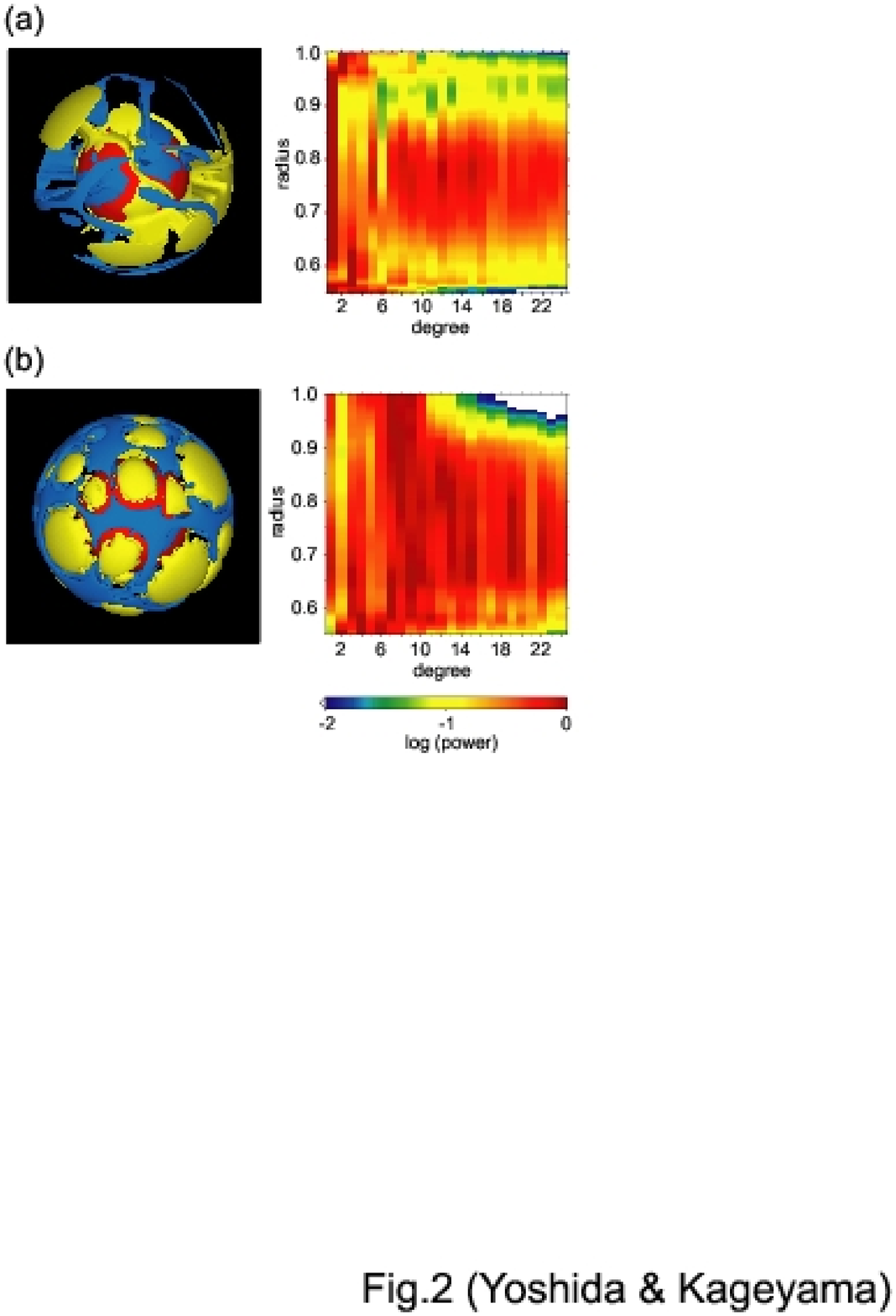}
\caption{
The iso-surface of residual temperature $\delta T$
(i.e., the deviation from horizontally averaged temperature at each depth),  
and the power spectrum of the spherical harmonics of temperature field at each depth.  
(a) Case~{\tt r7e0} with the constant viscosity (i.e., $E=0$) convection and 
(b) Case~{\tt r7e6r} with the strongly temperature-dependent viscosity ($E = \ln 10^6$) are shown. 
Blue iso-surfaces indicate $\delta T = $ (a) $-0.10$ and (b) $-0.15$.
Yellow indicate $\delta T = $ (a) $+0.10$ and (b) $+0.15$.  
The logarithmic power spectrum are normalized by the maximum values at each depth. 
White regions in maps indicate the values with lower than $10^{-2}$ (see color bars). 
}
\end{figure*}

\begin{figure}
\noindent\includegraphics[width=20pc]{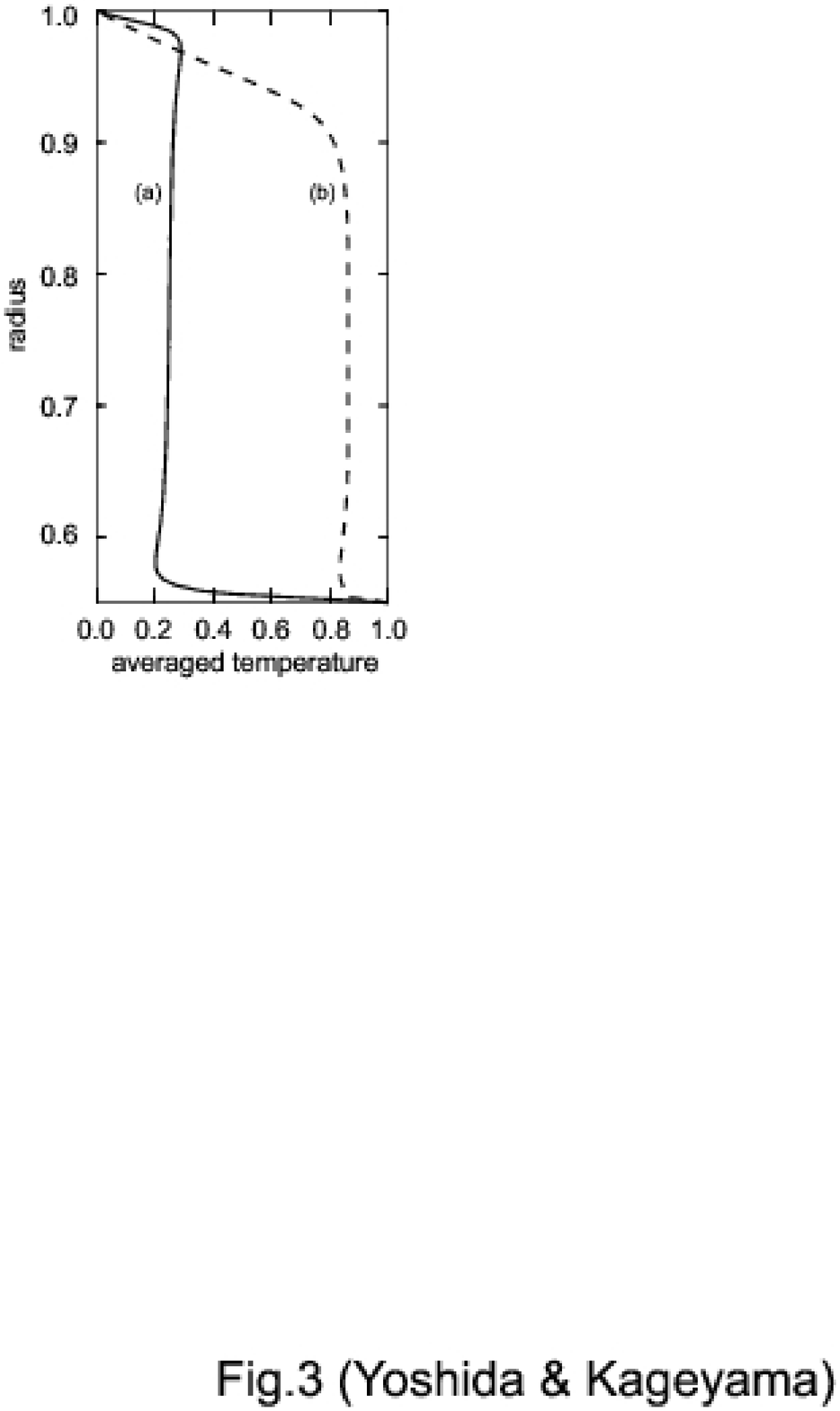}
\caption{
Radial profiles of horizontally averaged temperature at each depth. 
Two cases (a) and (b) correspond to each case shown in Figure 2 
(Cases ~{\tt r7e0} and {\tt r7e6r}, respectively). 
%The volume-averaged temperature is (a) 0.59, (b) 0.83, and (c) 0.72, respectively.
}
\end{figure}

\begin{figure*}
\noindent\includegraphics[width=30pc]{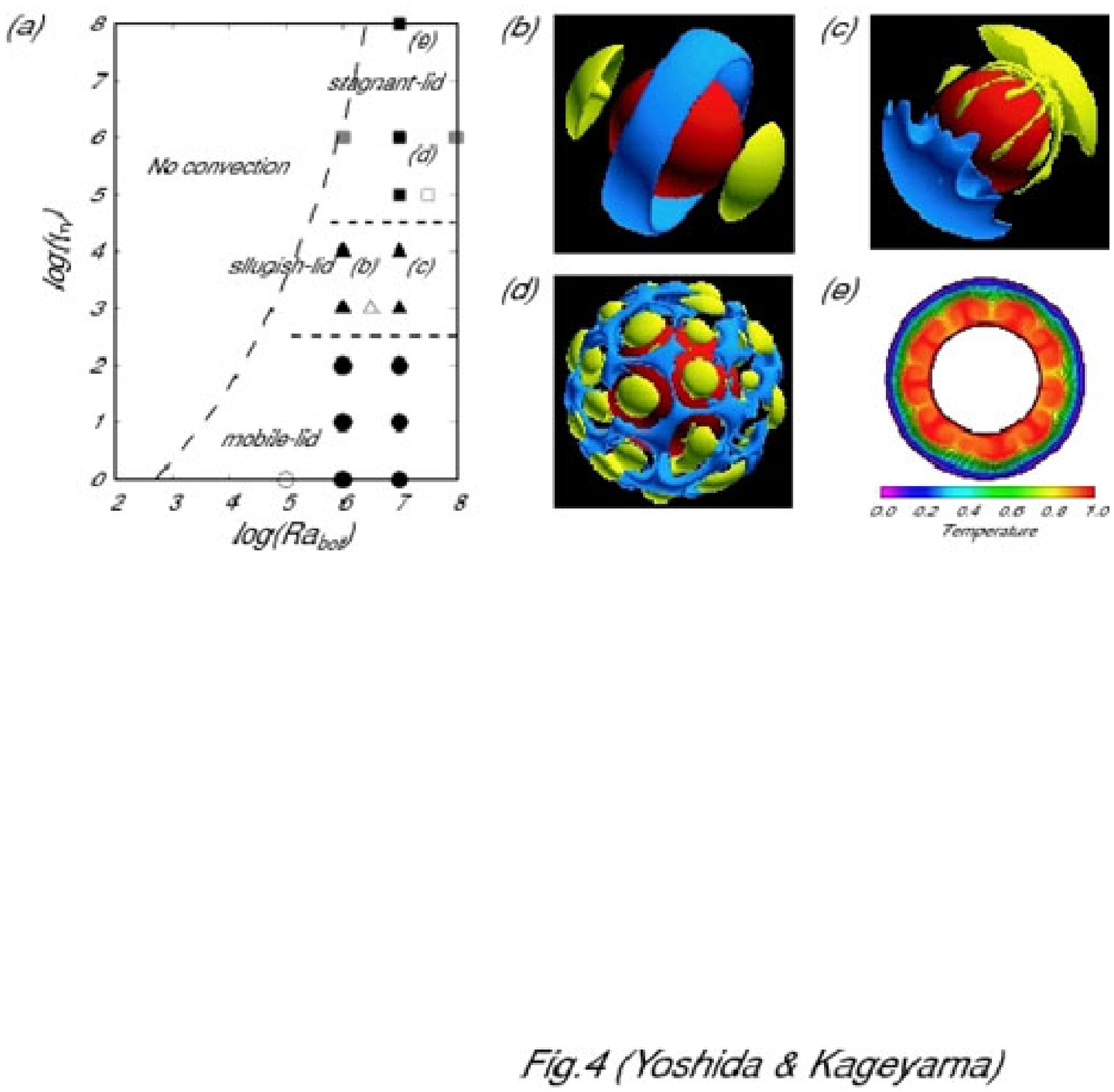}
\caption{
(a) The three convection regimes with varying Rayleigh number ($Ra_{bot}$) 
and the viscosity contrast across the shell ($\gamma_{\eta}$); 
the mobile-lid (circles), the sluggish-lid regime (triangles), and the stagnant-lid regimes (squares). 
Solid marks show our calculations. 
Open marks show the results from 3-D Cartesian box and spherical shell models 
by {\it Ratcliff et al.} [1997]. Gray marks show the results from 3-D Cartesian box models 
by {\it Trompart and Hansen} [1998]. 
The regime boundary (dashed curve) between convection regime and no-convection regime  
is referred with the reviews by {\it Schubert et al.} [2001]. 
Dashed line shows the approximate boundaries 
that separate the three convection regimes. 
(b)--(d) The iso-surface renderings of residual temperature 
shown in (a); 
(b) $Ra_{bot} = 10^6$ and $\gamma_{\eta} = 10^4$ (Case~{\tt r6e4}), 
(c) $Ra_{bot} = 10^7$ and $\gamma_{\eta} = 10^4$ (Case~{\tt r7e4}), and 
(d) $Ra_{bot} = 10^7$ and $\gamma_{\eta} = 10^6$ (Case~{\tt r7e6}). 
Blue iso-surfaces indicate (b) $\delta T = -0.20$, (c) $-0.25$, and (d) $-0.10$.
Yellow indicate (b) $\delta T = +0.40$, (c) $+0.25$, and (d) $+0.10$. 
The red spheres show the bottom boundary of the mantle.  
(e) The temperature distribution on a cross section for 
a case where $Ra_{bot} = 10^7$ and $\gamma_{\eta} = 10^8$ (Case~{\tt r7e8}).
}
\end{figure*}

\begin{figure}
\noindent\includegraphics[width=20pc]{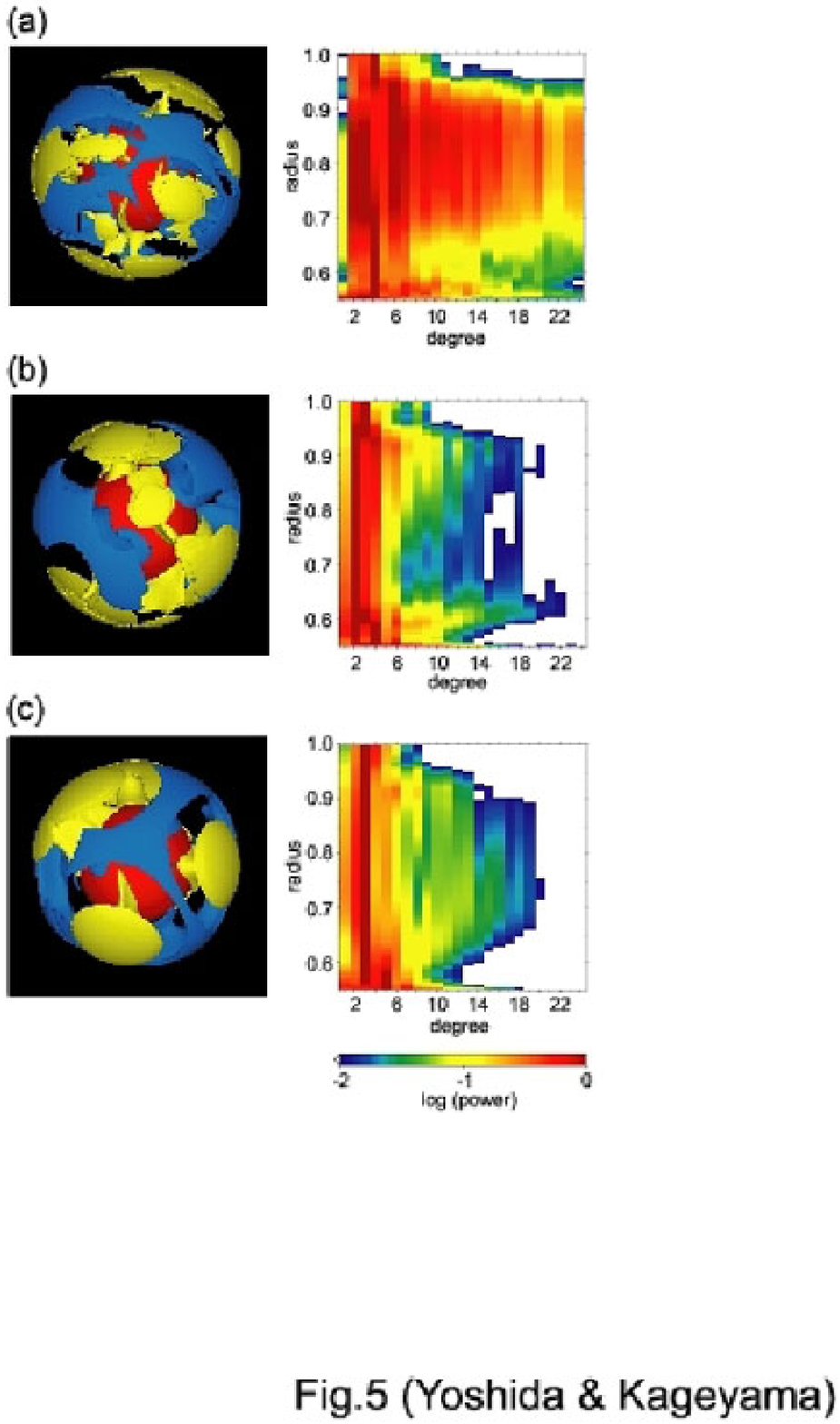}
\caption{
The iso-surface of residual temperature ($\delta T$)    
and the power spectrum of the spherical harmonics of temperature field at each depth 
for the cases where $\eta_{L}/\eta_{ref} =$ (a) $10^{1.5}$, (b) $10^{2.0}$, and (c) $10^{2.5}$ 
(Cases~{\tt r7e6v1}, {\tt r7e6v2}, and {\tt r7e6v3}, respectively). 
Blue iso-surfaces indicate (b) $\delta T = -0.20$, (c) $-0.25$, and (d) $-0.25$.
Yellow indicate (b) $\delta T = +0.20$, (c) $+0.25$, and (d) $+0.25$.  
The logarithmic power spectrum are normalized by the maximum values at each depth. 
White regions in maps indicate the values with lower than $10^{-2}$ (see color bars). 
}
\end{figure}

\begin{figure}
\noindent\includegraphics[width=20pc]{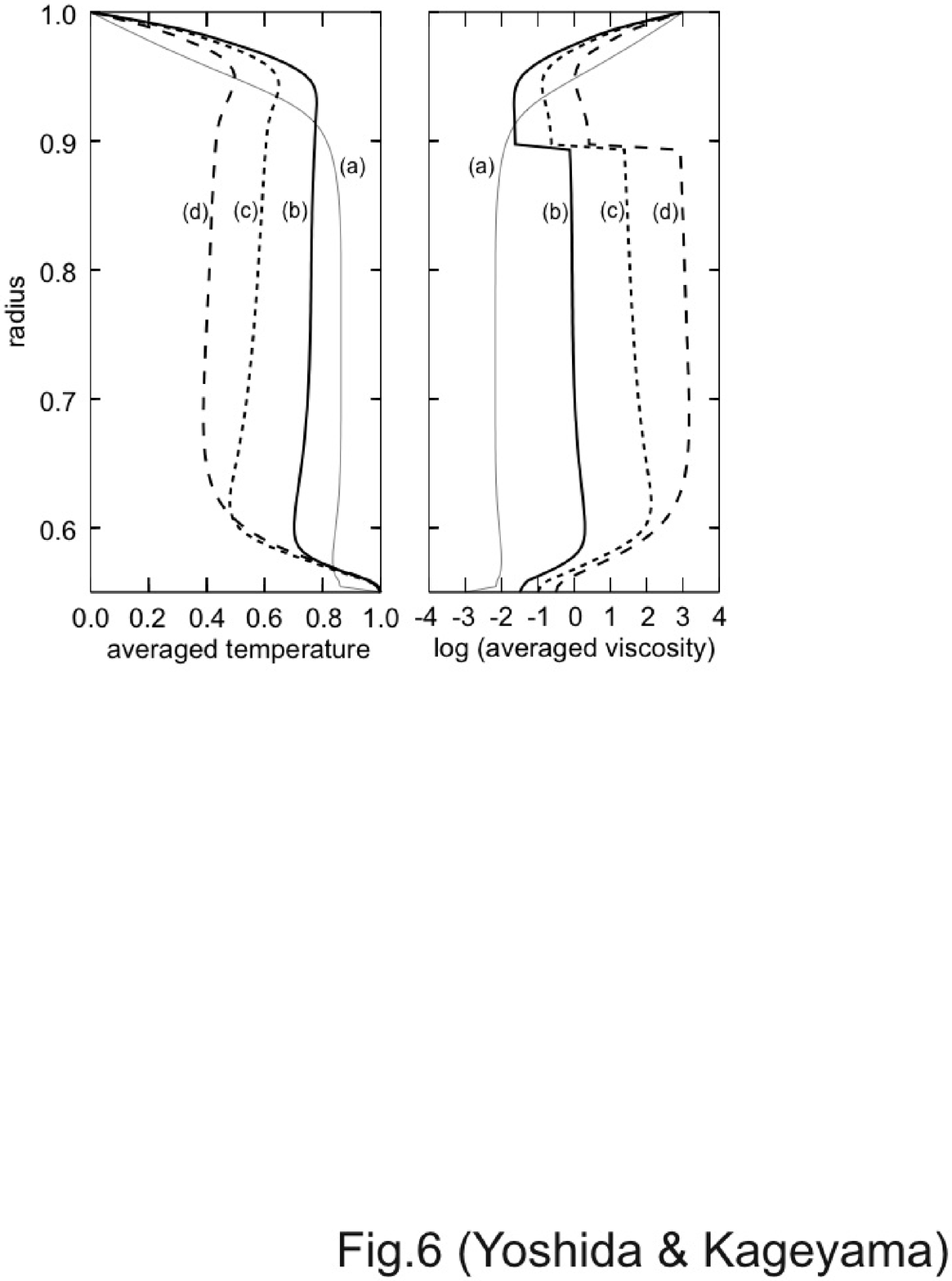}
\caption{
Radial profiles of the horizontally averaged temperature (left),  
and the horizontally averaged viscosity (right) at each depth. 
Three cases (a)--(d) correspond to each case  
where $\eta_{L}/\eta_{ref} =$ (a) $1$ (i.e., no viscosity stratification), (b) $10^{1.5}$, (c) $10^{2.0}$, and (d) $10^{2.5}$ 
(Cases~{\tt r7e6v1}, {\tt r7e6v2}, and {\tt r7e6v3}, respectively).  
%The volume-averaged temperature is (a) 0.71, (b) 0.56, and (c) 0.43, respectively.
}
\end{figure}

\begin{figure*}
\noindent\includegraphics[width=30pc]{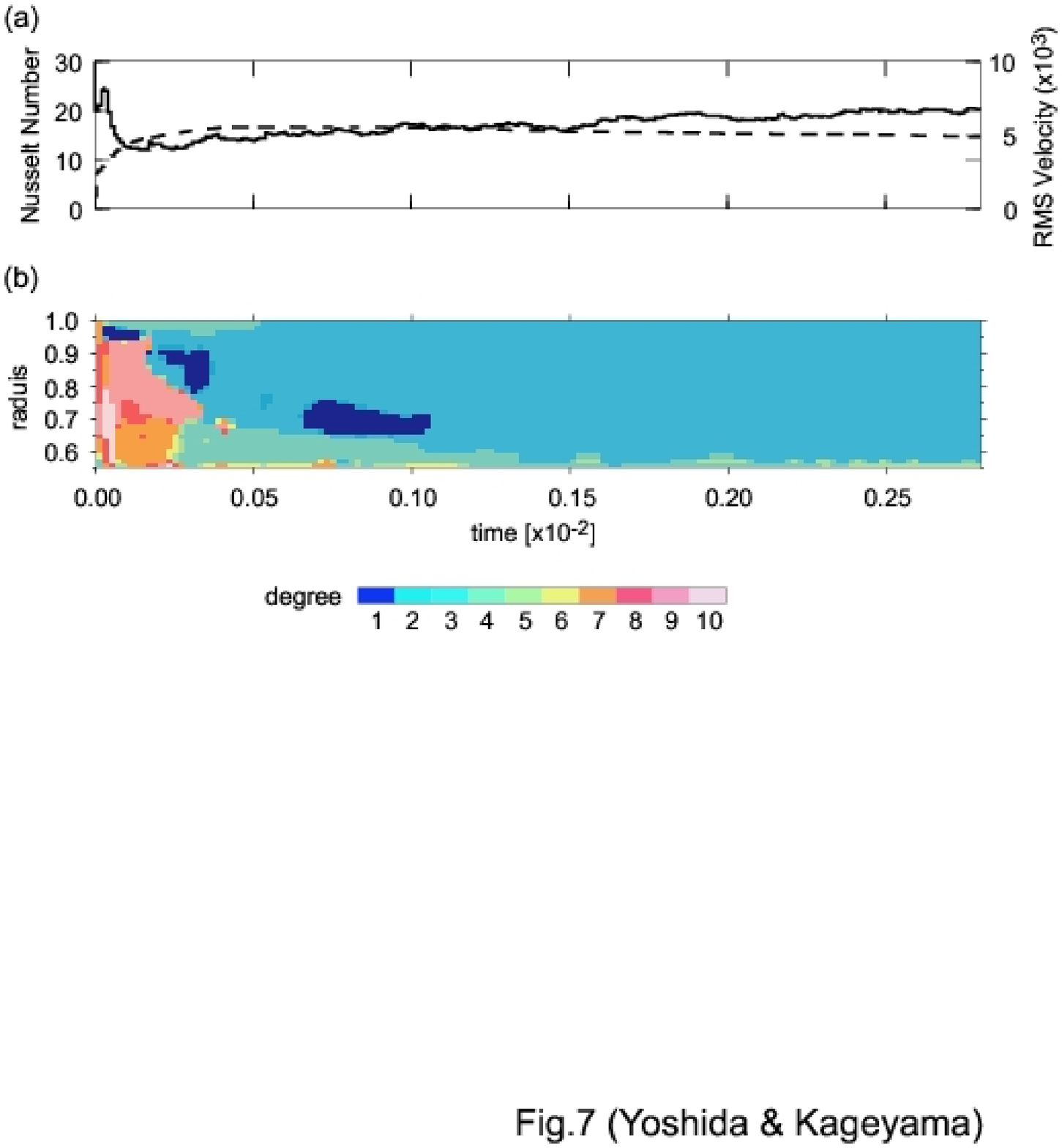}
\caption{
The time sequence of 
(a) the Nusselt number (dashed line) and the root-mean-square velocity averaged over the entire mantle (solid line), 
and (b) the maximum power spectrum at each depth. 
The range of the spherical harmonic degrees ($\ell$) is analyzed up to $\ell = 10$.
}
\end{figure*}

\begin{figure}
\noindent\includegraphics[width=18pc]{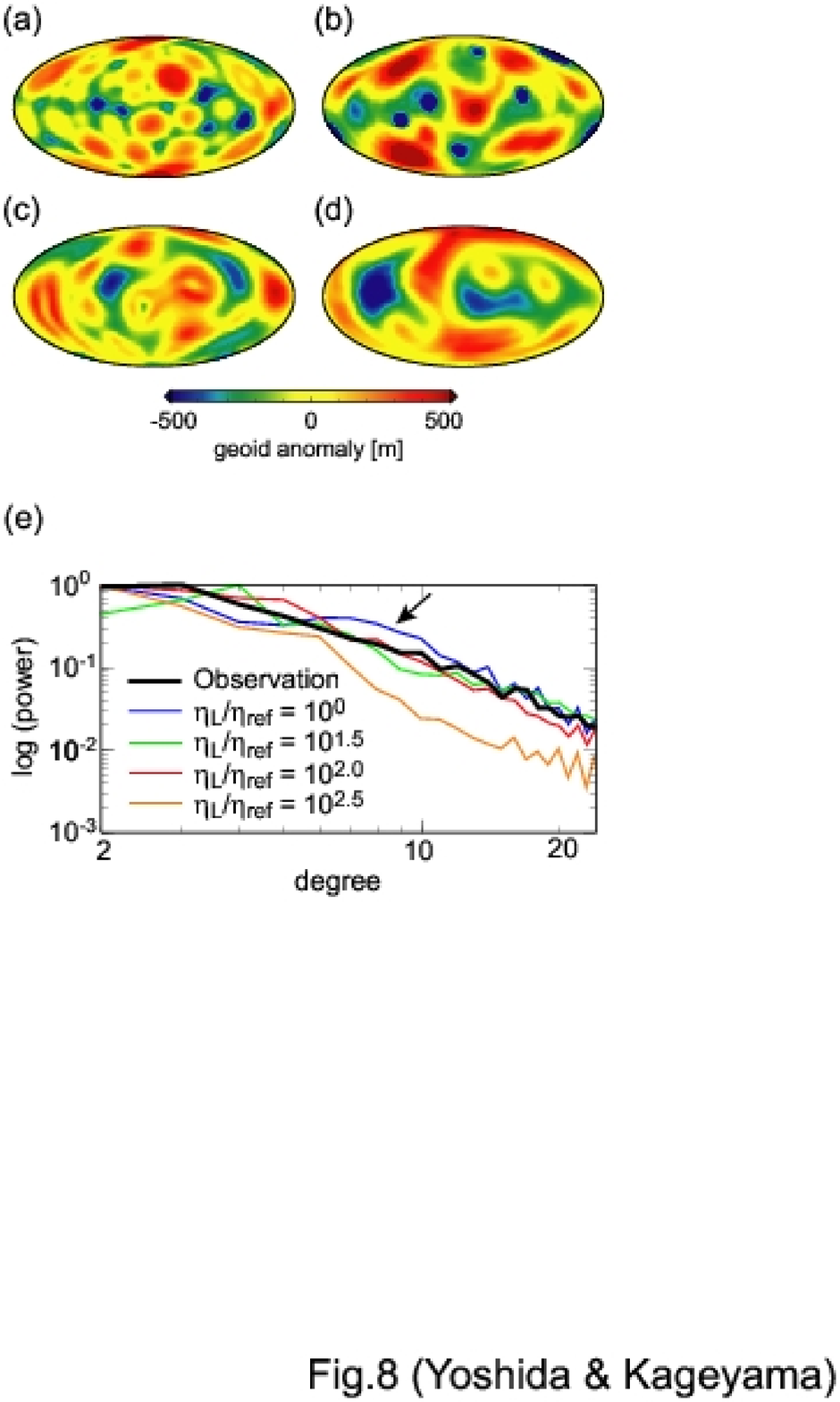}
\caption{
The contour plots of the distribution of geoid anomaly  
for each case where $\eta_{L}/\eta_{ref} =$ (a) 1 (i.e., no viscosity stratification), 
(b) $10^{1.5}$, (c) $10^{2.0}$, and (d) $10^{2.5}$. 
The results are shown by the spherical harmonic expansion up to $\ell = 24$. 
The spectrum are normalized by the maximum values at each degree. 
(e) The power spectrum of 
the calculated geoid anomaly for each case (thin colored lines) and  
the observed geoid anomaly from the data by {\it Konopliv et al}. [1999] (thick black line).  
The spectrum are normalized by the maximum values of all degrees. 
}
\end{figure}

\begin{figure}
\noindent\includegraphics[width=20pc]{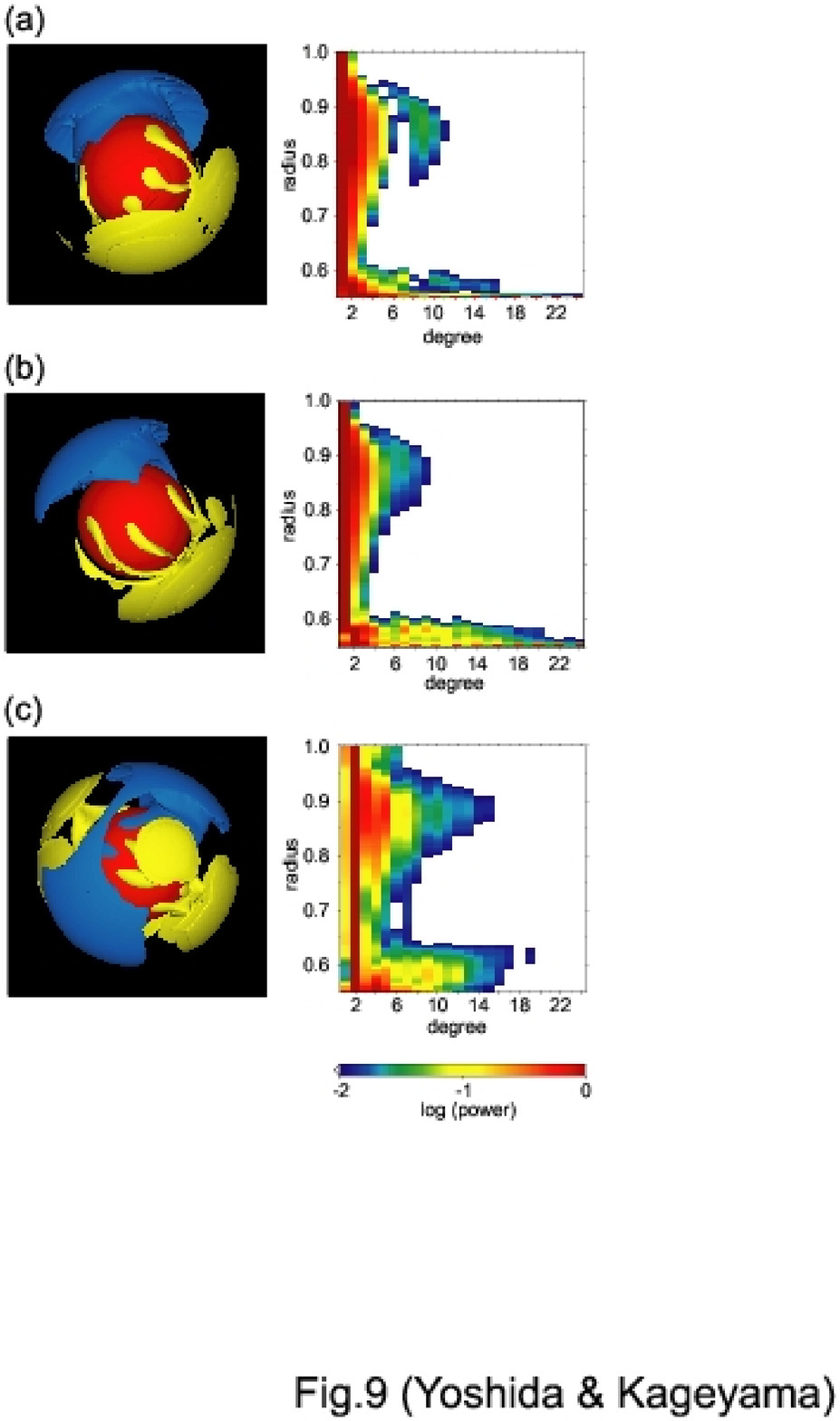}
\caption{
The iso-surface of residual temperature ($\delta T$)  
and the power spectrum of the spherical harmonics of temperature field at each depth 
for the cases where $\Delta \eta_L = $ (a) $10^{1.5}$, (b) $10^{2.0}$, and (c) $10^{2.5}$ 
(Cases~{\tt r7e6w1}, {\tt r7e6w2}, and {\tt r7e6w3}, respectively). 
Blue iso-surfaces indicate (a) $\delta T = -0.30$, (b) $-0.30$, and (c) $-0.20$.
Yellow indicate (a) $\delta T = +0.30$, (b) $+0.30$, and (c) $+0.20$.  
The logarithmic power spectrum are normalized by the maximum value at each depth. 
White regions in maps indicate the values with lower than $10^{-2}$ (see color bars). 
}
\end{figure}

\begin{figure}
\noindent\includegraphics[width=20pc]{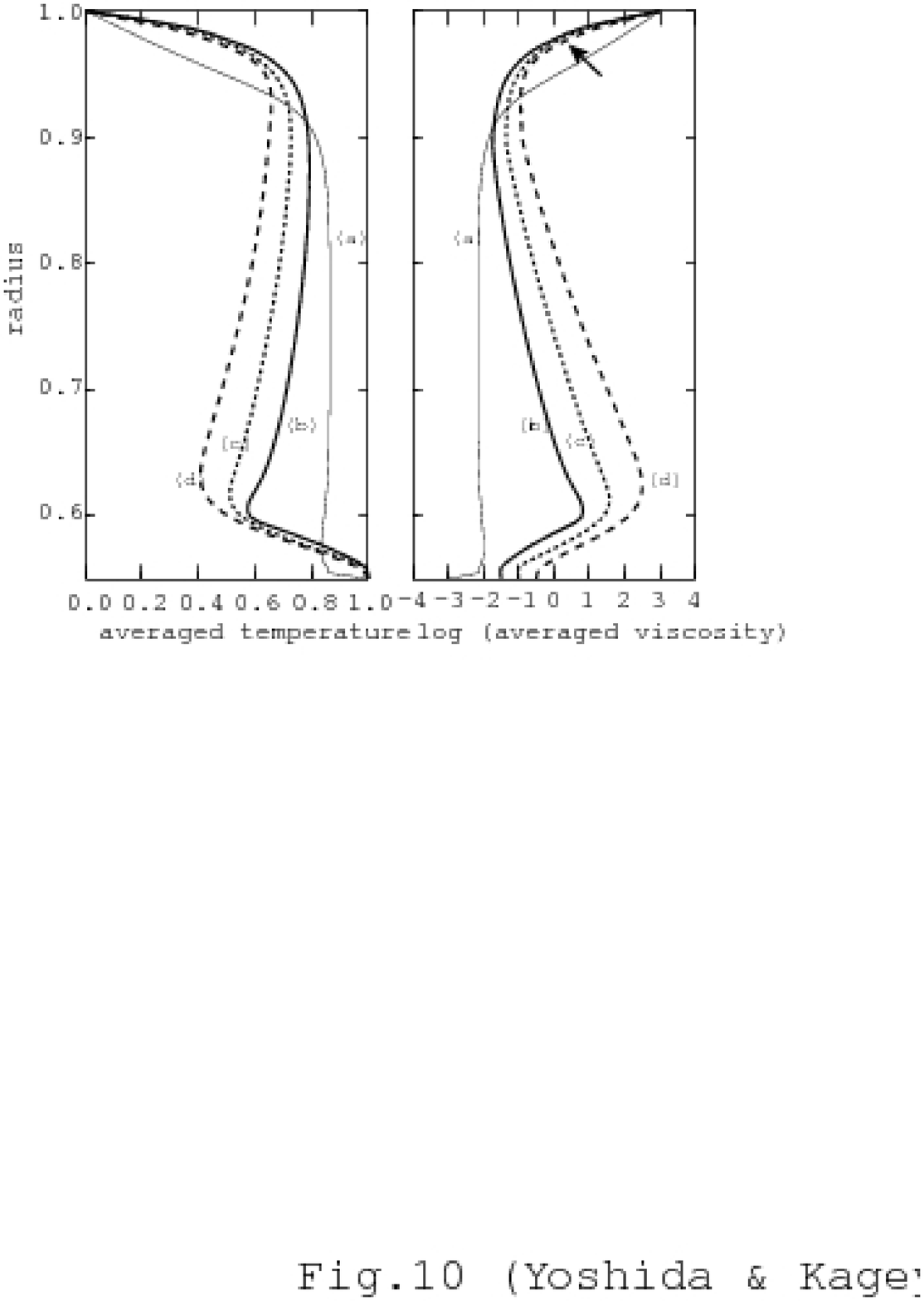}
\caption{
Radial profiles of horizontally averaged temperature (left),  
and the horizontally averaged viscosity (right) at each depth for each case 
where $\Delta \eta_L =$ (a) $1$ (i.e., no viscosity stratification), (b) $10^{1.5}$, (c) $10^{2.0}$, and (d) $10^{2.5}$ 
(Cases~{\tt r7e6w1}, {\tt r7e6w2}, and {\tt r7e6w3}, respectively). 
}
\end{figure}

\begin{table*}
 \caption{List of runs employed in this study}% \tablenotemark{a}}
 \label{symbols}
 \begin{tabular}{llllllll}
\hline
Case Name & $Ra$ & $T_{ref}$ & $E$ & $\eta_L/\eta_U$ & $\Delta \eta_L$ & I.C.$^{*1}$ & Corresponding figures \\
\hline
{\tt r6e0}   & $10^6$ & --  & $\ln 10^0$    & --             & --             & --   & -- \\
{\tt r6e1}   & $10^6$ & 1.0 & $\ln 10^1$    & --             & --             & {\tt r6e0} & -- \\
{\tt r6e2}   & $10^6$ & 1.0 & $\ln 10^2$    & --             & --             & {\tt r6e1} & -- \\
{\tt r6e3}   & $10^6$ & 1.0 & $\ln 10^3$    & --             & --             & {\tt r6e2} & -- \\
{\tt r6e4}   & $10^6$ & 1.0 & $\ln 10^4$    & --             & --             & {\tt r6e3} & Fig. 4b \\
{\tt r7e0}   & $10^7$ & --  & $\ln 10^0$    & --             & --             & {\tt r6e0} & Figs. 2a and 3 \\ 
%{\tt r7e0rg}$^{*2}$ & $10^7$ & --   & $\ln 10^0$    & --     & --                & {\tt r6e0} & Figs. 2b and 3 \\
{\tt r7e1}   & $10^7$ & 1.0 & $\ln 10^1$    & --             & --             & {\tt r7e0} & -- \\
{\tt r7e2}   & $10^7$ & 1.0 & $\ln 10^2$    & --             & --             & {\tt r7e1} & -- \\
{\tt r7e3}   & $10^7$ & 1.0 & $\ln 10^3$    & --             & --             & {\tt r7e2} & -- \\
{\tt r7e4}   & $10^7$ & 1.0 & $\ln 10^4$    & --             & --             & {\tt r7e3} & Fig. 4c \\ 
{\tt r7e5}   & $10^7$ & 1.0 & $\ln 10^5$    & --             & --             & {\tt r7e4} & -- \\
{\tt r7e6}   & $10^7$ & 1.0 & $\ln 10^6$    & --             & --             & {\tt r7e5} & Fig. 4d \\
{\tt r7e8}   & $10^7$ & 1.0 & $\ln 10^8$    & --             & --             & {\tt r7e6} & Fig. 4e \\
{\tt r7eA}   & $10^7$ & 1.0 & $\ln 10^{10}$ & --             & --             & {\tt r7e6} & -- \\ 
{\tt r7e6r}  & $10^7$ & 0.5 & $\ln 10^6$    & --             & --             & {\tt r7e6} & Figs. 2b and 3 \\
{\tt r7e6v1} & $10^7$ & 0.5 & $\ln 10^6$    & $\ln 10^{1.5}$ & --             & {\tt r7e6r} & Figs. 5a and 6 \\
{\tt r7e6v2} & $10^7$ & 0.5 & $\ln 10^6$    & $\ln 10^{2.0}$ & --             & {\tt r7e6r} & Figs. 5b and 6 \\
{\tt r7e6v3} & $10^7$ & 0.5 & $\ln 10^6$    & $\ln 10^{2.5}$ & --             & {\tt r7e6r} & Figs. 5c and 6 \\
{\tt r7e6w1} & $10^7$ & 0.5 & $\ln 10^6$    & --             & $\ln 10^{1.5}$ & {\tt r7e6r} & Figs. 9a and 10\\
{\tt r7e6w2} & $10^7$ & 0.5 & $\ln 10^6$    & --             & $\ln 10^{2.0}$ & {\tt r7e6r} & Figs. 9b and 10\\
{\tt r7e6w3} & $10^7$ & 0.5 & $\ln 10^6$    & --             & $\ln 10^{2.5}$ & {\tt r7e6r} & Figs. 9c and 10\\
{\tt r7e8w2} & $10^7$ & 0.5 & $\ln 10^8$    & --             & $\ln 10^{2.0}$ & {\tt r7e6w2} & -- \\
{\tt r7eAw2} & $10^7$ & 0.5 & $\ln 10^{10}$ & --             & $\ln 10^{2.0}$ & {\tt r7e6w2} & -- \\
{\tt r8e6w2} & $10^8$ & 0.5 & $\ln 10^6$    & --             & $\ln 10^{2.0}$ & {\tt r7e6w2} & -- \\
{\tt r7e6w2h}$^{*2}$ & $10^7$ & 0.5 & $\ln 10^6$   & --     & $\ln 10^{2.0}$ & {\tt r7e6w2} & -- \\
\hline
\end{tabular}
%\tablenotetext{a}{
\\
\tablenotetext{}{
(*1) ``I.C.'' indicates the Initial conditions.  
%(*2) Case {\tt r7e0rg} is a case with rigid boundary condition on the top surface boundary (see text). 
(*2) ``Case {\tt r7e6w2h}'' is a case with internal heating (see text).  
}
\end{table*}

\begin{table*}
 \caption{List of parameters used in the calculation of the geoid anomaly}
 \label{symbols}
 \begin{tabular}{lll}
\hline
                          & Symbols & Values \\
\hline
  outer radius            & $r_1$      & $6.052 \times 10^6$ m \\
  inner radius            & $r_0$      & $0.55 r_1$ m \\
  thickness of the mantle & $D$        & $0.45 r_1$ m \\
  density                 & $\rho$     & $3.3 \times 10^3$ kg m$^{-3}$ \\
  density contrast at the top surface    & $\Delta \rho_{bot}$ & $2.3 \times 10^3$ kg m$^{-3}$ \\
  density contrast at the bottom surface & $\Delta \rho_{top}$ & $4.3 \times 10^3$ kg m$^{-3}$ \\
  gravity acceleration    & $g$        & $8.9$ m s$^{-2}$ \\
  thermal expansivity     & $\alpha$   & $1.0 \times 10^{-5}$ K$^{-1}$ \\
  temperature difference across the mantle & $\Delta T$ & $2.0 \times 10^{3}$ K \\
  specific heat at constant pressure       & $c_p$      & $1.2 \times 10^3$ J kg$^{-1}$ K$^{-1}$ \\
  thermal diffusivity     & $\kappa = k / \rho c_p$   & $8.1 \times 10^{-7}$ m$^2$ s$^{-1}$ \\
  thermal conductivity    & $k$        & $3.2$ W m$^{-1}$  K$^{-1}$ \\
  reference viscosity     & $\eta$     & $1.5 \times 10^{21}$ Pa s \\
%%%  activation energy       & $E_a$      & $0-5.74  \times 10^4$ J mol$^{-1}$ \\
  gas constant            & $R$        & $8.3145$ J mol$^{-1}$ K$^{-1}$ \\
  gravitational constant  & $G$        & $6.6726 \times 10^{-11}$ N m$^2$ kg$^{-2}$ \\      
\hline
\end{tabular}
%\tablenotetext{a}{
\\
\tablenotetext{}{
The values are referred with 
{\it Schubert et al.} [1990], 
{\it Solomatov and Moresi} [1996], and 
{\it Turcotte and Schubert} [2002]. 
}
\end{table*}

\end{document}